**Search for Impact Craters in Iran: Citizen Science as a Useful Method**  H. Pourkhorsandi, Department of Geology, Faculty of Science, University of Tehran, Tehran 14155-64155, Iran (hkhorsandi@khayam.ut.ac.ir)

**Introduction:** Developments in remote sensing and satellite imagery since 60's have had a major role in the recognition of about 170 terrestrial impact structures. Free access to satellite images has led to the investigation of earth's surface by specialists and non-specialists, at the result these attempts have led to the discovery of new impact craters around the globe (e.g. [1-4]). Middle East as a vast region consists only three confirmed impact structures (Wabar craters [5], Kamil [1] and Jabal Waqf es Suwwan [2] craters). Compared to the other regions of the world very few researches on this topic have been done in the Middle East.

**Methods:** In an attempt to identify new impact craters in Iran, we formed a group consisted of volunteers from Iranian amateur astronomers society to investigate: 1- Satellite images via Google Earth© to identify "circular" structures that maybe be related to meteoritic impacts. 2- Reports in ancient books about the probable meteorite impact related phenomena. 3- Old reports and stories among rural people. Members got familiar with the topic with prepared web materials and lectures.

Central and Eastern Iran images were divided to 30*30 km quadrilaterals. Each member investigates a segment with a standard method and reports the circular structures. Then we investigate it with more images, topographic and geological maps and data of the region and we do a field study if it' necessary.

**Results:** Two examples of the investigated structures are reported here.

a) A circular structure with a diameter of ~ 200m situated ~100km NW of Birjand city identified (33°21'57"N, 58°14'24"E) (Fig. 1). This feature is not mentioned in 1:100000 geologic map of the area [6]. A field study took place in summer 2012. The area is composed of dark color sedimentary rocks. Bottom of the structure is composed of black limestone and black shale and the rim lithology is terrigenous sedimentary rocks with sharp and regular layering and some folded layers (N,S rims). There is no sign of brecciation and/or meteoritic fragments in the region that are primary diagnostic indicators for small size impact craters (e.g. [7]). So it is unlikely to consider it as an impact crater. It is probable that the structure has formed by activity of the small stream that passes inside crater and three faults that intersect each other in the place.

b) In his book "the Raiders of the Sarhad" General Dyer explains about a crater formed by a meteorite fall, based on local people said: "The old man had told him that, one night when he was a youth, something had exploded in the sky and fallen to the earth, punching a hole one hundred feet deep in the plain" [8]. There is a picture of this hole in [9](Fig. 2); based on crater shape it is unlikely to be an impact craters and is very similar to a sink hole.

This feature is not mentioned in maps. It is situated ~130km SW Zahedan (28°24'52" N, 60°34'44" E). In a field study in 2011, we find out that the crater is completely been filled with fine grained sediments and there is no evidence of a meteorite related phenomenon. Based on local's descriptions and the picture of the Gwarkuh crater it does not in the least suggest a meteorite crater ([10] also concluded the idea) and almost certainly it was a sink hole.

Beside these structures, field studies on another craters in Iran are in progress, the outcomes of which will be announced in the near future.

**References:**  [1] Folco L. et al. (2011) *Geology*, 39, 2, 179-182. [2] Salameh E. et al. (2008) *MAPS*, 43, 10, 1681-1690. [3] Glikson Y. et al. (2008) *AJES*, 55, 1107-1117. [4] Kofman R. S. et al. (2010) *MAPS*, 45, 9, 1429-1445. [5] Wynn J.C. and Shoemaker E.M. (1998) *SciAm*, 279, 5, 36-43. [6] *Geological Map of Sarghanj (1:100000).* (1995) GSI. [7] French B. M. and Koeberl C. (2010) *Earth-Sci. Rev.*, 98, 1-2, 123-170. [8] Dyer R. E. H. (1921) *the Raiders of the Sarhad*, H.F. & G. Whitherby, London, 268pp. [9] Skrine C.P. (1931) *GEOGR J*, 78, N, 321-340. [10] Spencer L. (1933) *GEOGR J*, 81, 3, 227-243.

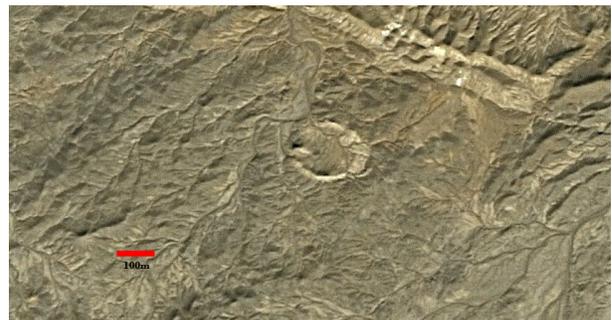

*Fig 1. Satellite image of a circular structure in Eastern Iran. (maps.google.com)*

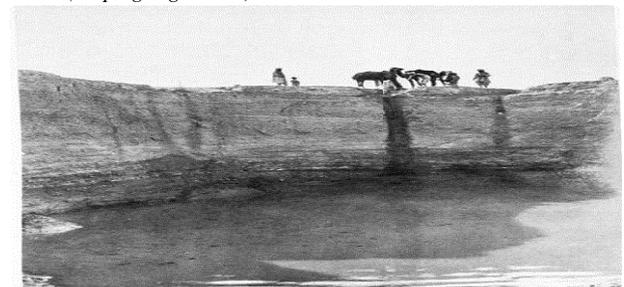

*Fig 2. Gwarkuh Crater. From Skirin C.P. [9].*